\documentclass[twocolumn, letter]{jpsj3}
\usepackage{graphicx}
\usepackage{dcolumn}
\usepackage{bm}
\usepackage{amsmath,amssymb}
\usepackage{color}
\newcommand{\nn}{\nonumber} 
 
\begin{document} 
 
\title{ 
Quantum Limiting Behaviors of a Vortex Core in an
Anisotropic Gap Superconductor
} 
\inst{Department of Physics, Okayama University,
Okayama 700-8530, Japan}  
\author{Shin-ichi \textsc{Kaneko}, Ken \textsc{Matsuba}, Muhammad \textsc{Hafiz}, Keigo \textsc{Yamasaki}, Erika \textsc{Kakizaki}, Nobuhiko \textsc{Nishida}, 
Hiroyuki \textsc{Takeya}$^{1}$, Kazuto \textsc{Hirata}$^{1}$, 
Takuto \textsc{Kawakami}$^{2}$\thanks{E-mail address: kawakami@mp.okayama-u.ac.jp}, Takeshi \textsc{Mizushima}$^{2}$, Kazushige  \textsc{Machida}$^{2}$ }
\inst{
Department of Physics, Tokyo Institute of Technology,
Tokyo 152-8551, Japan,

$^{1}$National Institute for Materials Science,
Tsukuba 305-0047, Japan,

$^{2}$Department of Physics, Okayama University,
Okayama 700-8530, Japan
} 
\date{\today}

\abst{ 
{Quantized} bound states at {a} vortex core {are discretized}
in YNi$_2$B$_2$C.
By using scanning tunneling spectroscopy with {an} unprecedented 0.1 nm spatial resolution, 
we find and identify the
localized spectral structure{,} where in addition to {the} first main peak
with a positive low energy, {a} second {subpeak} coming
from the {fourfold} symmetric gap structure is seen inside the energy gap. 
Those spectral features are understood
by solving the Bogoliubov-de Gennes equation for a {fully} three{-}dimensional
gap structure. {{A} particle-hole asymmetric spectrum}
at the core site
and quantum oscillation in the spectra {are}
clearly observed.}
\kword{{vortex bound states, STS, quantum limit, Bogoliubov-de Gennes equation, anisotropic superconductor}}

 

\maketitle 
{Q}uasiparticles (QP{s}) bound in a vortex core, such as Majorana particle{s},
are expected to play a fundamental role in various physical situations~\cite{nayak,kane},
ranging from {the} static and dynamical properties of vortex matter to
quantum computation where it is required to exchange QPs bound in a core to encode a qubit
through non-Abelian statistics\cite{ivanov,nayak}.
In order to isolate and manipulate those QPs, we need to identify {such} particles
experimentally and theoretically as a first step.
Namely{,} we need to find the quantized energy levels for QPs {using}
{spectroscopic methods}, such as scanning tunneling microscopy and spectroscopy (STM/STS).
This task is still {far from being realized} although several STM/STS experiments have probed
the broad peak centered {at the approximately} zero-bias energy $E=0$ associated with the
vortex bound states~\cite{hess,nishimori}, namely{,} {the} so-called Caroli-de Gennes-Matricon (CdGM) state~\cite{caroli}, but those experiments {did not} resolve each discretized energy level fully quantum-mechanically.
The origin of the difficulty in observing a discretized core bound state is due to {the fact that}
(A) $T$ should be {sufficiently low} to satisfy the condition $T<T_c^2/\epsilon_F$ ($T_c$ is the superconducting
transition temperature and $\epsilon_F$ is the Fermi energy).
(B) The {spatial} resolution of STM/STS should be {sufficiently fine} to see the isolated bound state.
Those two conditions have not been fulfilled in the exiting STM/STS experiments.

In our previous STS stud{y} of the vortex bound state in YNi$_2$B$_2$C at 0.46K \cite{nishimori},
we have observed broad peaks with electron-hole asymmetry in intensity in the tunneling spectrum 
around the center of the vortex core and 
interpreted them in terms of a quantum regime vortex core where the wave function of the {hole-like} bound state
starts from zero at the center and that of the electron-like bound state does from {the} maximum. 
In that {study,} the spatial resolution was about 1 nm and the {average} spectra in the {2$\times$2 nm$^2$} of the vortex core were measured.
Here,  we have performed {an} STM/STS experiment at lower $T$ 
with {an} unprecedented spatial resolution ($\sim$0.1 nm), expecting to clarify {the} quantum nature of 
vortex core states and {to} possibl{y} observ{e} new features such as {the} quantum oscillations of QPs
{in} YNi$_2$B$_2$C where $T_c$=15.6 K, $H_{c2}$=8.8 T, the mean 
free path $l$=32 nm, the coherent length $\xi$=6 nm{,} and the
penetration depth $\lambda$=110 nm.
We {have also} theoretically analyzed {the obtained spectra}
to reveal that the STS data
indeed exhibit crucial information on discretized quantum level structure{s}.

A {homemade} scanning tunneling microscope {can} be operated at temperature
down to 180 mK and {at} magnetic fields of up to 6 T.
We used {a} Pt-Ir alloy wire with a diameter of 0.3 mm that {was} mechanically sharpened
in this experiment. 
Before cooling down the sample, we checked the tip condition over the Au film.
STS measurement {was} conducted simultaneously with STM measurement.
Using STM, we scan{ned} the sample for a wide surface where the square 
lattice can be seen. We {chose} a suitable spot and size range for the STS
measurement. Bias voltage and feedback current {were} set before 
{the} measurement. We measured the $I$-$V$ characteristic{s} of the sample and 
{numerically} calculated {the} differentiation of the $I$-$V$ characteristic{s}
to find the differential conductance. 
We normalize{d} $dI/dV$
with the slope of the $I$-$V$ characteristic{s} by fitting it 
linearly. Then the vortex {was} imaged by plotting the value
of $dI/dV$ at various biases.

In YNi$_2$B$_2$C{,} a variety of experiments~\cite{izawa,park,baba} 
{have shown} on the existence of point nodes, 
such as angle-resolved specific heat~\cite{park} and thermal conductivity~\cite{izawa}.
From {an} STS experiment, the tunneling spectrum was found to have {an} $E^3$ dependence
near $E=0$ in the superconducting gap {at} zero magnetic field. In order to explain
the $E^3$ dependence, the existence of point nodes with the asymmetric recovery
and the need {for} 3D calculation were proposed~\cite{nishimori2}.
Angle{-}resolved photoemission measurement shows {a} strong gap magnitude variation on
the cylindrical Fermi surface in the 17th {band} at the X point of the tetragonal Brillouin zone
of this material~\cite{baba}.
It is known that {a} two{-}dimensional $d$-wave nodal gap 
does not yield {an} isolated vortex bound state inside the bulk gap because all the
low{-}energy wave functions are extended, leaking out from
the nodal directions~\cite{franz},
in contrast to the isotropic $s$-wave gap~\cite{hayashi1}.
Thus{,} it is not {clear whether} the point node case {exhibits} the
CdGM like bound state.

Our analytic {strategy is to solve} the Bogoliubov-de Gennes {(BdG)} equation in three dimensions (3D)
for various possible 3D gap functions, including the point node gap function
on a cylindrical Fermi surface.
The BdG formalism deals with the quantum limiting regime 
beyond the quasiclassical approximation~\cite{ichioka, nagai1, nagai2, hayashi2} 
in the sense of the accessibility to discretized {eigenstates} 
and their quantum oscillation with {the order of lattice spacing}.

{In general, the BdG equation in spin singlet superconducting states is given as} 
\begin{eqnarray}\label{generalbdg}
	\int d\bm{r}_2
	\left[\begin{array}{lr}
	H_0(\bm{r}_1,\bm{r}_2) & \Delta(\bm{r}_1,\bm{r}_2) \\
	\Delta^*(\bm{r}_1,\bm{r}_2) & -H_0^*(\bm{r}_1,\bm{r}_2)
	\end{array} \right]\vec{u}_\nu(\bm{r}_2)\! \nn\\
	=\! E_\nu\vec{u}_\nu(\bm{r}_1),
\end{eqnarray}
where the single particle Hamiltonian 
$H_0\!=\!\delta(\bm{r}_1\!-\!\bm{r}_2)\left(-\frac{\hbar^2\nabla^2_1}{2m}\!+\!V(\bm{r}_1)\!-\!\epsilon_F\right)$ 
and the {QP} wave function $\vec{u}_\nu(\bm{r})\!=\![u_\nu(\bm{r}),\ v_\nu(\bm{r})]^T$ 
{with the Fermi energy $\epsilon_F$}. 
The pair potential $\Delta(\bm{r}_1,\bm{r}_2)$ is decomposed into $\Delta(\bm{R})$ 
in the center-of-mass coordinate $\bm{R}\!=\!(\bm{\rho}_1\!+\!\bm{\rho}_2)/2$,
{where $\bm{\rho}_i\!=\!\bm{x}_i\!+\!\bm{y}_i$,}
and the gap function $\varphi(\tilde{\bm r})$ in the relative coordinate
$\tilde{\bm r}\!=\!\bm{r}_1\!-\!\bm{r}_2$.
Here, we consider the single vortex state described as $\Delta(\bm{R})=\Delta_0\tanh(|\bm{R}|/\xi)\exp(i\phi)$, 
where the coherence length $\xi\!=\!\epsilon_F/(2\Delta_0 k_F)$ 
and the azimuthal angle $\phi\!=\!\arctan[(x_1+x_2)/(y_1+y_2)]$.
In this model, the wave number $k_z$ becomes a {well-defined} quantum number 
because of the translation symmetry along the $\bm{z}${-}axis.
The BdG equation (\ref{generalbdg}) is reduced to the $k_z$-resolved {two-dimensional} form
\begin{eqnarray}\label{kzrbdg}
	\int d\bm{\rho}_{{2}}
	\left[\begin{array}{cc}
	H_0(\bm{\rho}_1,\bm{\rho}_2) & \Delta(\bm{R})\varphi_{k_z}(\tilde{\bm \rho}) \\
	\Delta^*(\bm{R})\varphi_{k_z}(\tilde{\bm \rho}) & -H_0^*(\bm{\rho}_1,\bm{\rho}_2)
	\end{array} \right]\vec{u}_{\nu,k_z}(\bm{\rho}_{{2}})\! \nn\\ 
	=\! E_{\nu, k_z}\vec{u}_{\nu,k_z}(\bm{\rho}_{{1}}),
\end{eqnarray}
where $H_0\!=\!\delta(\bm{\rho}_1\!-\!\bm{\rho}_2)\left(-\frac{\hbar^2\nabla'^2_1}{2m}\!+\!V(\bm{\rho}_1)\!-\!\epsilon_F^{{2D}}(k_z)\right)$ 
and $\nabla'^2\!=\!\partial_x^2\!+\!\partial_y^2$. 
The two-dimensional form of the Fermi energy $\epsilon_F^{2D}(k_z)\!=\!\frac{\hbar^2}{2m}(k_F^2-k_z^2)$ 
reflects the $k_z$-cross section of the Fermi surface.
{The index $\nu\!\in\!\mathbb{Z}$ denotes the $\nu$-th excited state 
of the $k_z$-resolved BdG equation, {eq.}~(\ref{kzrbdg}).}
The gap function $\varphi_{k_z}(\bm{\rho})\!=\!\frac{1}{(2\pi)^2}\int dk_x dk_y\varphi_{\bm{k}} e^{i\bm{k}\cdot{\bm \rho}}$ 
is obtained from $\varphi_{\bm k}$ in the ${\bm k}$-space by performing the fast Fourier transform.
We can solve the BdG equation for the {arbitrary} anisotropic function $\varphi_{\bm k}$.
Then, we obtain the QP local density of states (LDOS)
\begin{eqnarray*}
	\mathcal{N}(\bm{\rho},E)&\!=\!\sum\limits_{k_z}\mathcal{N}_{k_z}(\bm{\rho},E)\!=\!\sum\limits_{\nu,k_z}|u_{\nu,k_z}(\bm{\rho})|^2\delta_{\eta}(E\!-\!E_{\nu,k_z}),
\end{eqnarray*}
with the Lorentzian function $\delta_\eta(z)\!=\!\eta^2/[z^2\!+\!\eta^2]$.
$\eta$ is adjusted to compare the experimental data of the STS measurements below.
{The procedure of the numerical {diagonalization} of the BdG equation is explained in {ref}.~\ref{mizushima}}.


\begin{figure}[tb]
\begin{center}
\includegraphics[width=6.0cm]{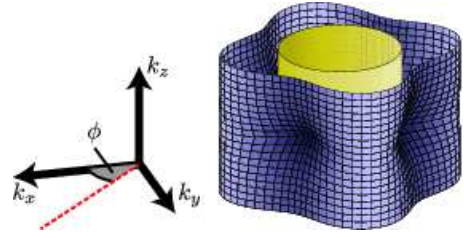} 
 \caption{ 
(Color online) Schematic view of the gap function and definition of the angle $\phi$
drawn on the cylindrical Fermi surface used in the calculation.
} \label{fig:DFT}
\end{center} 
\end{figure} 

It is considered that the gap function {of YNi$_2$B$_2$C} is suppressed at four points, 
where {nesting vectors exist}~\cite{maki, nagai2}.
We parametrize the gap function {as}
\begin{eqnarray}
\varphi_{\hat{\bm k}}=\frac{1}{2}\left\{1+\varphi_\mathrm{min}(k_z)+[1-\varphi_\mathrm{min}(k_z)]\cos 4\phi_k\right\}{,}
\end{eqnarray}
where {the angle} $\phi_k\!=\!\arctan(k_x/k_y)$ {and} {in-plane gap minimum} 
$\varphi_\mathrm{min}(k_z)\!=\!\varphi_0[\tanh(k_z/\alpha k_1)]^n$.
This gap function {is equivalent to the $s\!+\!g$-wave gap~\cite{maki} function 
in the $k_z\!=\!0$ plane{,} and} {the} parameters $\alpha$, $n$, {and} $\varphi_0$ 
signify how {rapidly} the point nodes at the {$k_x\!=\!0$ and $k_y\!=\!0$} plane 
recover the full gap with increasing $k_z$ toward the Brillouin zone boundary at $k_z\!=\!k_1$.
We choose the cylindrical Fermi surface as {the} first approximation, where $\epsilon^{2D}_F(k_z)$ is constant.
This is because we focus on the 17th Fermi surface around {the} X point of YNi$_2$B$_2$C, which does not close 
in the Brillouin zone as obtained by {the} band calculation~\cite{yamauchi}.
Figure~\ref{fig:DFT} shows a three{-}dimensional view of $\varphi_{\hat{\bm k}}$ 
along with the cylindrical Fermi surface. 
As we will see later, the combination of the Fermi surface shape and the
gap function along the $k_z$ direction is crucial {for} reproduc{ing} the STS data.
{We carry out the calculation within the quantum limit regime $k_F\xi\!=\!10$.}
{In the case of the YNi$_2$B$_2$C, the direction of the Fermi velocity $\bm{v}_F$ {at} gap nodes~\cite{yamauchi}}
{is different by $\pi/4$ from that of the wave vector $\bm{k}_F$,}
{so that the low-energy QP {spectrum rotates} by $\pi/4$}
{{from} the isotropic case $\bm{v}_F\!\parallel\!\bm{k}_F$~\cite{nagai1}.}
{Thus, the spectrum along the [110] ([100]){-}axis in our experiments corresponds}
{to that along the $\phi=0\ (\pi/4)$ in our calculations.}

\begin{figure}[tb] 
\includegraphics[width=8.5cm]{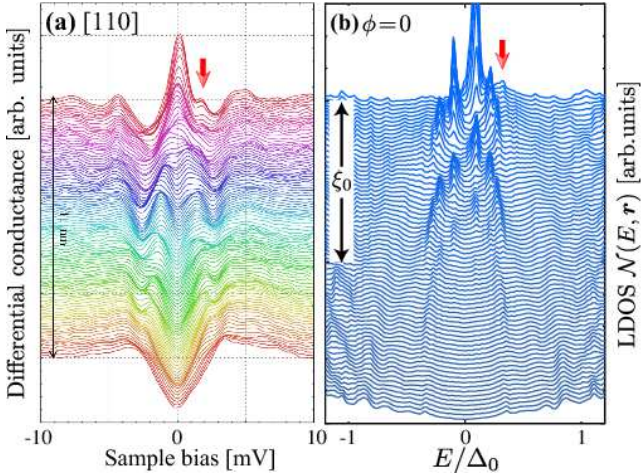} 
 \caption{ 
(Color online) 
{(a) {Differential} conductance at $T$=190mK and $H$=0.3T as a function of the sample bias obtained by STS measurements for [110] direction}. {The energy resolution is about 0.1 meV in this measurment.} The spacing between {spectra} is 0.11 nm.
The total path length is 11 nm. 
(b) The spectral evolutions of the LDOS $\mathcal{N}(E, {\bf r})$ for {antinodal} direction. 
The spatial scale is denoted by long arrows
with {a} length $\xi_0$. The short down arrows in the back
indicate the shoulder structure at the core.
It is seen that several low energy peaks near $E=0$
oscillate and evolve in their intensities as a function of the position
from the core.
}\label{fig:ARDOS}
\end{figure}

Figure~\ref{fig:ARDOS} shows the experimental (a) and theoretical (b) spectral evolution{s} 
along {{the} direction{s} (a) [110] and (b) $\phi\!=\!0$}.
{In the experimental data shown in Fig.~\ref{fig:ARDOS}(a), it is evident to see {that}}
(i) {the main peak at the core is situated {on} the positive energy side of about 0.3 meV with {an} asymmetric shape.}
(ii) The {2nd peak} at the core site {exists at} the positive bias $E\!\sim\!1.5$ meV, 
{marked with the arrow in Fig.~\ref{fig:ARDOS}(a)}.
(iii) The spatial modulated behavior appears on the evolution ridge.
We note that the average of the present STS data with {a} 0.1 nm spatial resolution in {a} 2$\times$2 nm{$^2$}
well reproduce{s} the previous STS results of tunneling spectra by Nishimori $et$ $al.$~\cite{nishimori}
{These features (i)-(iii) are supported by the numerical calculation shown in Fig.~\ref{fig:ARDOS}(b).}
In particular, in Fig.~\ref{fig:ARDOS}(b), the feature (iii) is interpreted 
as the quantum oscillations of the peak intensities {associated with} several lower energy excitations.
We also find that (iv) quantum oscillation depends 
on either {the} {{$\phi\!=\!0$ or $\pi/4$}} direction.
The finite energy spectrum near $E/\Delta_0\!\sim\!0.3$ in the {$\phi\!=\!0$} direction 
extend longer than that in the {$\phi\!=\!\pi/4$} direction.
This angular dependence is also shown in the experimental data 
as the difference between the evolution of the [110] and [100] direction{s}.
These {consistencies} with the theory denote that the experimental data in Fig.~\ref{fig:ARDOS} 
{shows} the quantum limiting behaviors {of} the vortex core bound states
{whose discreteness is specified by the energy scale $\Delta_0^2/\epsilon_F^{2D}$.}
It is also found that {the spherical or ellipsoidal Fermi surface {smoothens} the quantum oscillation of the LDOS.}
{Hence, the cylindrical Fermi surface is necessary for the quantum oscillation.}

\begin{figure}[tb] 
\includegraphics[width=8.5cm]{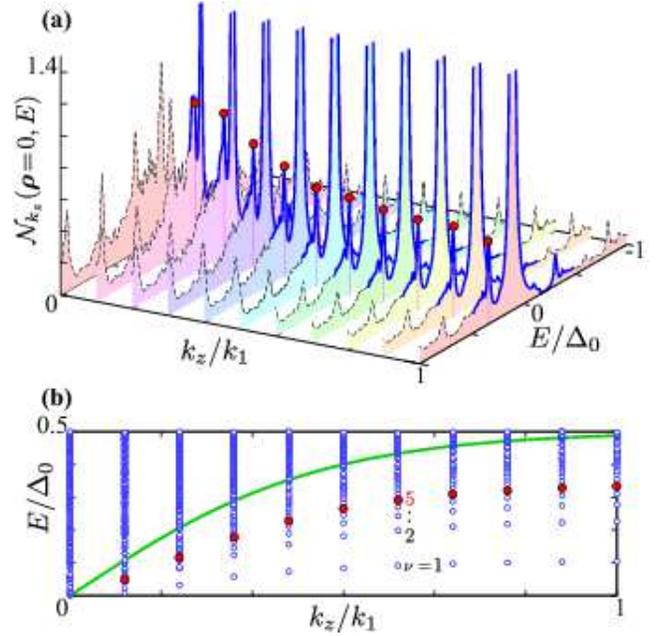} 
 \caption{ 
(Color online) 
(a) The $k_z$-resolved spectra $\mathcal{N}_{k_z}(r\!=\!0,E)$ at the core.
{The solid (broken) line indicates inside (outside) of the in-plane gap minimum $\Delta_0\varphi_\mathrm{min}$.}
It is seen that as $k_z$ increases the 2nd peak structure
at $E/\Delta_0=0.3$ becomes split from  the main park at $E=0.1$.
(b) The $k_z$-resolved eigenenergy $E_{\nu,k_z}$ (symbols) and the gap minimum $\Delta_0\varphi_\mathrm{min}$ (line).
Filled symbols in the both panels denote the eigenenergy $E_{\nu\!=\!5,k_z}$, the 5th excited state of the $k_z$ resolved BdG equation (\ref{kzrbdg}).
}
\label{fig:KZRDOS} 
\end{figure}

It is emphasized that the 2nd peak {that} appears at $E/\Delta_0\!\sim\!0.3${,} {shown} in Fig.~\ref{fig:ARDOS}(b), 
is a consequence of the rotational symmetry due to the point nodes 
and {is} never seen in the LDOS of isotropic $s$-wave superconductors~\cite{hayashi1}.
In order to understand the physical origin of the 2nd peak, we decompose the spectrum into the $k_z$-{resolved}
spectra as shown in Fig.~\ref{fig:KZRDOS}{(a)}{,} where the  $k_z$-resolved spectra at the core are {shown}.
The strong {peak} at $E/\Delta_0\!=\!{0.1}$ {is associated with} the lowest bound state $\nu\!=\!1$.
This $\nu\!=\!1$ peak becomes stronger {and shifts to the positive side} 
as $k_z$ increases because the gap becomes wider.
Simultaneously{,} the 2nd peak at  $E/\Delta_0\!=\!0.3$ grows and is {distinctly} separated from the main
$\nu\!=\!1$ peak  {when $k_z\!\neq\!0$}.
Figure~\ref{fig:KZRDOS}(b) shows the $k_z$-resolved eigenenergy {of the BdG equation {eq.}~(\ref{kzrbdg})}.
At $k_z\!=\!0$, where the in-plane gap minimum {is} {$\varphi_\mathrm{min}(k_z\!=\!0)\!=\!0$}, 
the spectrum is {continuous}, and the eigenstates with {$E/\Delta_0$} lower than 
the gap minimum {$\varphi_\mathrm{min}$} become discrete with increasing $\varphi_\mathrm{min}$.
{The} $\nu\!=\!5$ {eigenenergy} is marked in Fig.~\ref{fig:KZRDOS} with filled symbols, 
which {consists of the 2nd peak in the LDOS}.

{The physical origin is explained as follows.}
As $\varphi_\mathrm{min}\!\rightarrow\!1$, 
{which is} {the} limit of {a} two-dimensional isotropic $s$-wave case~\cite{hayashi1}, 
the branches of the eigenstates $\nu$ approach the eigenstates quantized 
by the angular momentum $\ell\!=\!1/2,\ 3/2,\ 5/2,\cdots$ with the energy 
${E}_{\ell\!+\!\frac{1}{2},k_z}\!=\!\ell\omega_0$, where $\omega_0\!\sim\!\Delta_0^2/2\epsilon_F^{2D}$, 
and the wave function 
$u_{\ell\!+\!\frac{1}{2},k_z}(\bm{\rho})\!\propto\!J_{{\!\ell\!-\!\frac{1}{2}}}(k_F\rho)$ 
near the vortex core, {where $J_{\nu}(x)$ is the Bessel function~\cite{caroli, hayashi1}}.
The only $\ell\!=\!1/2$ state has {a} nonzero core-site wave function 
$u_{\ell\!+\!\frac{1}{2},{k_z}}(\bm{\rho}\!=\!0)\!\neq\!0$.
When $\varphi_\mathrm{min}\!<\!1$, {however,} the {fourfold} rotationally symmetric pair potential {hybridizes}
with the eigenstates {with} the angular momentum $\ell\!+\!4n$ ($n\!\in\!\mathbb{Z}$)~\cite{ichioka}.
Thus, the $\ell\!=\!1/2$ state can be hybridized to the $\ell\!=\!1/2\!+\!4n$ states, 
{which are realized as the $\nu\!=\!4n\!+\!1$ states described as}
\begin{eqnarray}
	u_{\!4n\!+\!1}(\bm{\rho})=\sum\limits_{n'}a_n^{n'}J_{4n'}(k_F\rho){e^{-\!i4n\phi}},
\end{eqnarray}
This $n'\!=\!0$ contribution {of} the $\ell=1/2$ state 
in the wave function $u_{\nu\!=\!5}$ {is the} origin {of} the 2nd peak of the core-site LDOS.
The {peak} arising from the $\nu\!=\!5$ state is most enhanced 
because the $\ell=9/2$ state is the nearest{-}energy state from the $\ell\!=\!1/2$ state.
{Although there {are} {peaks} from the $\nu\!=\!-3,\ 9\ \cdots$ states in the $k_z$-resolved core-site LDOS,}
{these peaks cannot be seen in Fig.~\ref{fig:ARDOS}(a) {or \ref{fig:ARDOS}(b)} with the $k_z$ integration.}

\begin{figure}[t] 
\includegraphics[width=8.5cm]{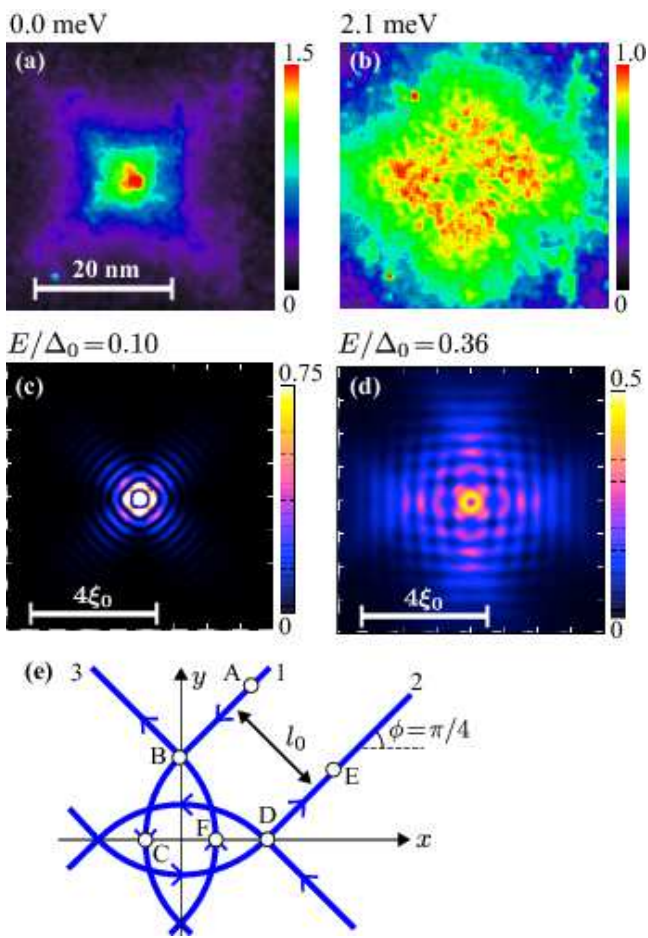} 
 \caption{ 
(Color online) Experimental differential conductance $dI/dV$ with the sample biases $E\!=\!0.0$ meV (a) and $E\!=\!2.1$ meV (b) at $T$=180 mK and $H$=0.3 T.
The corresponding two-dimensional LDOS map $\mathcal{N}(E\!=\!0.1\Delta_0, {\bm \rho})$ (c) and $\mathcal{N}(E\!=\!0.36\Delta_0, {\bm \rho})$. The panel (e) shows the schematic figure of the QP paths with the large contribution when the vortex core exists at the origin.
} 
\label{fig:LDOSmap} 
\end{figure} 

Next, we move to the two-dimensional {profile} of the LDOS.
Figure~\ref{fig:LDOSmap} shows that the experimental and theoretical two-dimensional LDOS{s} 
with different {energies}.
It is known that the LDOS has {a} long tail along {the nodal direction $\phi\!=\!\pi/4$} near $E\!=\!0${;} 
{as the energy increases}{,} {the {direction} of the tails} {rotates} to {$\phi\!=\!0$}
in the case of the {fourfold} rotationally symmetric gap function at low {energies}.
{The corresponding theoretical LDOS is shown in Figs.~\ref{fig:LDOSmap}(c) and \ref{fig:LDOSmap}(d)}
{where clear two-dimensional quantum oscillations are {observed}.}
{Although in Figs.~\ref{fig:LDOSmap}(a) and \ref{fig:LDOSmap}(b) those two-dimensional oscillations}
{are hardly visible {owing to the} lack of spatial resolution in this case,}
{the quantum oscillations are well anticipated from Fig.~\ref{fig:ARDOS}(a) where the linear line cut}
{of the two-dimensional LDOS with the finest spatial resolution is displayed.}
{On the other hand, the depleted spectral weight on the vortex center and the surrounding four peaks}
{observed in Fig.~\ref{fig:LDOSmap}(b) are not reproduced in the theoretical LDOS} 
{in Fig.~\ref{fig:LDOSmap}(d), because of the technical difficulty {in} the numerical diagonalization.}

The two-dimensional oscillatory {behaviors are} understandable with the schematic picture of the QP path 
shown in Fig.~\ref{fig:LDOSmap}(e).
In the {fourfold} rotationally symmetric gap {function}, 
the {low{-}energy} QPs move from {the} {$\phi\!=\!\pi/4$} direction 
to the $\phi\!=\!-\!\pi/4$ direction 
around the vortex core~\cite{ichioka,nagai1,nagai2,hayashi2}.
The interference between the wave functions of the countercurrent QPs 
{leads to} {a} standing wave, {that is,} the oscillation of the LDOS.
When the energy {is} $E\!\sim\!0$ and the distance is $l_0\!\ll\!k_F^{-1}$ in Fig.~\ref{fig:LDOSmap}(e), 
the QP {``1''} moving from A to B and {the QP} {``2''}  moving from D to E interfere {with} each other.
The resulting {wavefront} direction of the standing wave is {$\phi\!=\!\pm\pi/4$}{,} 
as shown in Fig.~\ref{fig:LDOSmap}(c).
On the other hand, when the energy $E$ becomes {larger}, 
the distance between paths becomes {as} large as $\l_0\!\gtrsim\!k_F^{-1}$ with the growth of the impact parameter. 
The large $l_0$ allows {the} {formation of} the interference pattern 
by the QP {``1''} moving from B to C and {the QP} {``3''} moving from F to B in Fig.~\ref{fig:LDOSmap}(e).
Thus, the LDOS oscillation pattern {appears along} the {$\phi\!=\!0,$ $\pi/2$} direction 
in the high{-}energy region $E/\Delta_0\!=\!0.36${,} as shown in Fig.~\ref{fig:LDOSmap}(d).
{The change {in} the orientation of the LDOS is clearly observed in {the} {experimental results} in Fig.~\ref{fig:LDOSmap}}.
{{The observation of} the rotation of the oscillation pattern remains a future problem.}


We have performed STM/STS experiments
on YNi$_2$B$_2$C under {an} unprecedented spatial resolution of 
$\sim$0.1 nm at low temperatures.
Through the full quantum mechanical analysis based on 
the Bogoliubov-de Gennes equation for three{-}dimensional gap
structure{s}, 
we have succeeded in identifying at the core position
the quantized bound state at a positive energy as the
first peak with the particle-hole asymmetric shape. We have also found the second peak induced 
by the {fourfold} symmetric gap structure.
This is the first experimental observation of {clearly discretized Caroli-de Gennes-Matricon}
{vortex bound states}.
We conclude that{,} in reciprocal space{,} the point node positions correspond to the continuum spectra with {a} zero energy{,}
while the gapped positions give rise to discrete spectra.
{Note that the energy of first peak at the vortex core center $\delta E\!\sim\!0.3 $ meV in Fig.~\ref{fig:ARDOS}(a) is much larger than $\Delta_0^2/\epsilon_F$, where the Fermi energy is $\epsilon_F\!\sim\!1$ eV. However, if we estimate that as $\delta E\!=\Delta_0/k_F\xi_0$, where the experimentally obtained value is $k_F\xi_0\!\sim\!10$, the resulting $\delta E$ is reasonable. One of the reasons for the discrepancy between these estimates is that the realistic dispersion relations of YNi$_2$B$_2$C differ from the free electron model used in our analysis. In order to achieve quantitative estimates, one should take into account the realistic dispersion relations.}

We are grateful to M. Ichioka for stimulating discussion{s}.
This work was supported in part by {the} 21st Century G-COE Program {of} Tokyo {T}ech "Nano Science and Quantum Physics",
JSPS (No.~2200247703, 2074023303, 2134010303), and the ``Topological Quantum Phenomena" (No.~22103005) and the "Heavy Electrons" (NO.~20102006) KAKENHI {in} innovation areas from MEXT.


\end{document}